\documentclass[%
reprint,
amsmath,amssymb,
aps,
pre,
]{revtex4-2}

\usepackage{graphicx}% Include figure files
\usepackage{dcolumn}% Align table columns on decimal point
\usepackage{bm}% bold math
\usepackage[hidelinks]{hyperref}% add hypertext capabilities

\usepackage{amsmath, amsbsy}
\usepackage{amssymb}

\newcommand{\D}{\mathrm{d}}

\newcommand{\WG}{\Delta W_\text{G}}

\newcommand{\WD}{\Delta W_\text{D}}

\begin{document}
	
%	\preprint{APS/123-QED}
	
\title{Minimum Stabilizing Energy Release for Mixing Processes}	
\date{\today}
\author{E. J. Kolmes}
\email{ekolmes@princeton.edu}
\author{N. J. Fisch}
\affiliation{Department of Astrophysical Sciences, Princeton University, Princeton, New Jersey, USA}

\begin{abstract}
Diffusive operations, which mix the populations of different elements of phase space, can irreversibly transform a given initial state into any of a spectrum of different states from which no further energy can be extracted through diffusive operations. 
We call these ground states. 
The lower bound of accessible ground state energies represents the maximal possible release of energy. 
This lower bound, sometimes called the diffusively accessible free energy, is of interest in theories of instabilities and wave-particle interactions. 
On the other hand, the upper bound of accessible ground state energies has escaped identification as a problem of interest. 
Yet, as demonstrated here, in the case of a continuous system, it is precisely this upper bound that corresponds to the paradigmatic ``quasilinear plateau" ground state of the bump-on-tail distribution. 
Although for general discrete systems the complexity of calculating the upper bound grows rapidly with the number of states, using techniques adapted from treatments of the lower bound, the upper bound can in fact be computed directly for the three-state discrete system. 
\end{abstract}
 
\maketitle

\section{Introduction}

It is often of interest to calculate how much kinetic energy could be released from a given system -- that is, the system's free or available energy. 
There are multiple definitions of free or available energy, each corresponding to a different rule for how a distribution of particles may be rearranged. One of the simplest, due to Gardner in 1963 \cite{Gardner1963}, is that any rearrangement is permitted so long as it conserves phase space densities. 
These rearrangement operations are known as ``Gardner restacking." 
The maximum energy that can be extracted with Gardner restacking is known as the ``Gardner free energy."

However, physical processes that conserve phase space densities on a microscopic scale can appear to produce diffusion when the system is viewed with finite granularity, which is often the case of practical interest. 
For example, wave-particle interactions are often modeled as diffusive processes.  This includes the well-known collisionless damping mechanism of waves in plasma known as ``Landau damping" \cite{Landau1946}, where the damping of the wave is accompanied by diffusion of particles in velocity space, often modeled by quasilinear diffusion \cite{Kennel1966}. 
The diffusion of particles by waves underlies mechanisms of heating plasma by waves \cite{Stix} and mechanisms for driving plasma currents by waves \cite{Fisch1987}. 
In the event that the plasma is out of equilibrium, the diffusion of particles by waves can result in the amplification of the waves. 
Since the amplification is limited by how much energy can be released in a diffusive process, it can be useful to define the free energy limited by diffusive exchange, in which the allowed operation is to average the populations of any two elements of phase space (as opposed to Gardner restacking, where the populations instead exchange position without mixing) \cite{Fisch1993, Hay2015, Hay2017, Kolmes2020ConstrainedDiffusion, Kolmes2020Gardner}. 
These mixing operations are perhaps the simplest class of operations that do not conserve phase space densities. 

\begin{figure}
	\centering
	\includegraphics[width=.9\linewidth]{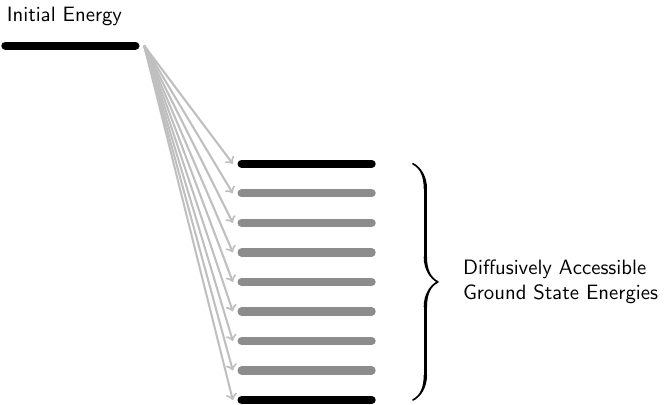}
	\caption{Diffusive operations can often map a given initial state into any of a large number of different ground states. These accessible ground states can have a range of energies. }
	\label{fig:spectrumCartoon}
\end{figure}

In this context, a ground state is defined as a state from which no operation can release further energy. 
For both Gardner restacking and diffusive exchange, the ground state is always a state in which the highest-population elements of phase space occupy the lowest-energy states. 
For any given initial state, Gardner restacking can lead to only one possible ground state, whereas diffusive exchange operations can lead to a spectrum of ground states (as is drawn in Figure~\ref{fig:spectrumCartoon}). 
The diffusively accessible free energy is defined as the energy released when the system is transformed from its initial state to the lowest-energy ground state that can be reached through diffusive operations. 
Calculating this free energy is therefore a search problem over the space of all accessible ground states. 

The diffusively accessible free energy was originally defined in the context of alpha channeling, where waves are intentionally injected into a system in order to extract energy from a population of fusion products \cite{Fisch1992, Fisch1993}. 
The motivation was to determine how efficient alpha channeling (and similar strategies for intentional phase space manipulations) could possibly be. 
This helps to explain why the focus in the diffusive-exchange literature has always been on the \textit{lowest}-energy ground state: for the purposes of engineering phase-space transformations to release as much energy as possible, the upper limit on the achievable energy release is the most interesting thing to calculate. 

The Gardner restacking literature, on the other hand, is largely motivated by the physics of instabilities. 
This includes Gardner's original work \cite{Gardner1963} as well as much of the recent progress on the theory of restacking \cite{Helander2017ii, Helander2020, Mackenbach2022}. 
If an instability can be understood as drawing energy from the unstable configuration, then the amount of energy that could possibly be extracted quantifies how unstable the system could be, without recourse to the dynamics of the particular instabilities in question. 
Recent work suggests that the Gardner free energy can sometimes provide powerful predictions for turbulent energy fluxes \cite{Mackenbach2022}. 

This difference in focus is largely historical rather than having anything to do with the underlying physics of these different transformations. 
However, if one wishes to use the theory of diffusive exchange operations to understand instabilities, then it becomes desirable to understand the rest of the spectrum of ground states pictured in Figure~\ref{fig:spectrumCartoon}, not just the lowest-energy state. 
After all, a natural instability will not necessarily pick the optimal sequence of phase space mixing operations; in general it may drive the system to any of the accessible ground states. 
This paper takes the first step toward understanding the rest of that spectrum by introducing the concept of the minimum stabilizing energy release -- that is, the identification of the \textit{highest}-energy ground state that can be reached through mixing operations. 
For an experimentalist hoping to avoid detrimental instabilities, this represents the best-case scenario: the smallest energy release that can stabilize the system. 
More importantly, when taken together with the (maximum) diffusively accessible free energy, the minimum stabilizing energy release quantifies the range of possible outcomes that can be achieved through mixing operations. 

These formulations of the available-energy problems come from the plasma physics literature, and are connected with a number of other ideas about stability and accessibility within that literature \cite{Taylor1963, Taylor1964, Morrison1989, Morrison1998}. 
However, these considerations are much more broadly relevant. 
Gardner restacking is closely related to ideas that appear in astrophysics \cite{Berk1970, Bartholomew1971, Lemou2012, Chavanis2012}, statistical mechanics \cite{Baldovin2016}, and mathematics \cite{Riesz1930, HardyLittlewoodPolya, Brascamp1974, Almgren1989, Baernstein}. 
The discrete diffusive exchange problem can be found (under other names) in the literature on physical chemistry \cite{Zylka1985}, income inequality \cite{Dalton1920, Atkinson1970, Aboudi2010}, and altruism \cite{Thon2004}. 
All of these formulations generally approach the problem of determining the set of states that can be reached under a particular set of operations. 
This more general problem appears, for example, in meteorology \cite{Lorenz1955}; chemistry \cite{Horn1964}; laser absorption \cite{Levy2014}; and quantum information theory and thermodynamics \cite{Gorban, Gorban2013, Lostaglio2015, Brandao2015, Korzekwa2019, Lostaglio2022}. 

This paper is organized as follows. 
Section~\ref{sec:definition} defines the minimum stabilizing energy release for discrete and continuous phase spaces. 
Section~\ref{sec:threeBox} explicitly calculates the minimum stabilizing release for a three-state discrete system, and describes how the problem differs from that of calculating the maximum energy release (that is, the minimume-energy accessible ground state). 
Section~\ref{sec:continuous} discusses the minimum stabilizing energy release for continuous phase space. 
It shows that the quasilinear plateau is the maximum-energy accessible ground state for a bump-on-tail distribution, and that this theory provides a natural generalization of the quasilinear plateau for more general curves. 
Section~\ref{sec:conclusion} discusses these results. 
Appendix~\ref{appendix:threeBox} describes explicit solutions for the minimum-energy accessible ground states for two- and three-state discrete systems, as well as the corresponding Gardner restacking ground states. 

\section{Defining the Minimum Stabilizing Energy Release} \label{sec:definition}

All of the aforementioned concepts of available energy can be defined for either discrete or continuous phase space. 
For the purposes of building intuition, it is often helpful to start with the discrete case. 
One can think of a discrete phase space as being a coarse-grained average over a continuous space. 
Alternatively, one can think of a discrete phase space as corresponding to a fundamentally discrete system (such as an atomic system with some discrete set of energy levels). 

A discrete system with $N$ states is specified by the energies $\{ \varepsilon_i \}$ and current populations $\{ n_i \}$ of those states; the total energy can be written as 
\begin{gather}
W = \sum_{i=1}^N \varepsilon_i n_i. 
\end{gather}
It is convenient to assume without loss of generality that $\epsilon_i \leq \epsilon_j$ $\forall i < j$, so that the system is in a ground state if and only if $n_i \geq n_j$ $\forall i < j$. 

A Gardner restacking operation consists of exchanging $n_i$ and $n_j$. 
A diffusive exchange operation consists of sending both $n_i$ and $n_j$ to $(n_i + n_j) / 2$. 
In the original formulation of the minimum-energy ground state problem, there was no further restriction on the allowed operations. 
However, when considering the maximum-energy ground state problem, it is necessary also to impose that an averaging operation should only be allowed if it does not increase the total energy. 
The disallowed operations, which effectively inject energy into the system, are sometimes called annealing operations. 
Annealing operations must be prohibited because, if they are allowed, the problem becomes both trivial and unphysical. 

It becomes trivial because the solution is always the same: every element of phase space is repeatedly averaged against every other element until all populations are equal. 
This outcome is unphysical; in the limit of large $N$, and in the continuous limit, it can involve an arbitrarily large increase in energy. 
In the continuous analog (which is described more fully below), this would correspond to a uniform distribution over the entire domain of velocity space. 
Moreover, these annealing operations are intrinsically not in line with how we typically expect instabilities to behave. 
Annealing operations were not prohibited in the original formulation of the minimum-energy ground state problem, but Hay, Schiff, and Fisch showed \cite{Hay2015} that the minimum-energy accessible ground state is the same with or without these operations. 

For a continuous phase space, the corresponding free energies can largely be understood in terms of the large-$N$ limit of the discrete problem. 
In the case of Gardner restacking, the continuous problem \cite{Dodin2005} is equivalent to calculating the ``symmetric decreasing rearrangement" discussed in the mathematics literature \cite{Riesz1930, HardyLittlewoodPolya, Brascamp1974, Almgren1989, Baernstein}. 
The continuous diffusive problem can be presented as an optimization problem on the energy 
\begin{gather}
W_\text{final} = \lim_{t \rightarrow \infty} \int \varepsilon(v) f(v,t) \, \D v
\end{gather}
for a distribution $f(v,t)$ that evolves in time through the non-local mixing process 
\begin{gather}
\frac{\partial f}{\partial t} = \int K(v,v',t) \big[ f(v',t) - f(v,t) \big] \, \D v'. 
\end{gather}
There is no requirement that the mixing be local because microscopically local flows can result in non-local mixing on larger scales \cite{Fisch1993}. 
The minimum-energy ground state problem is to find the kernel $K(v,v',t)$ that minimizes $W_\text{final}$, with the requirements that $K(v,v',t) = K(v',v,t)$ and $K(v,v',t) \geq 0$. 
The maximum-energy ground state problem is to instead maximize $W_\text{final}$, with the added constraints that $K(v,v',t)$ can only be nonzero when $\varepsilon(v) - \varepsilon(v')$ and $f(v,t) - f(v',t)$ have the same sign (no annealing) and that the final state must be a ground state. 

The space of allowed kernels $K(v,v',t)$ is large, and direct searches over this space are difficult. 
However, it was recently shown (surprisingly enough) that the minimum-energy diffusively accessible ground state energy for a continuous system is identical to the energy accessible through Gardner restacking \cite{Kolmes2020Gardner}. 
This is to be contrasted with discrete systems, in which the Gardner free energy always exceeds the energy accessible through diffusive exchange (with the exception of the case in which the system starts in a ground state and there is no free energy of either kind). 

\section{$N=3$ Discrete Case} \label{sec:threeBox}

\begin{figure*}
	\centering
	\includegraphics[width=.48\linewidth]{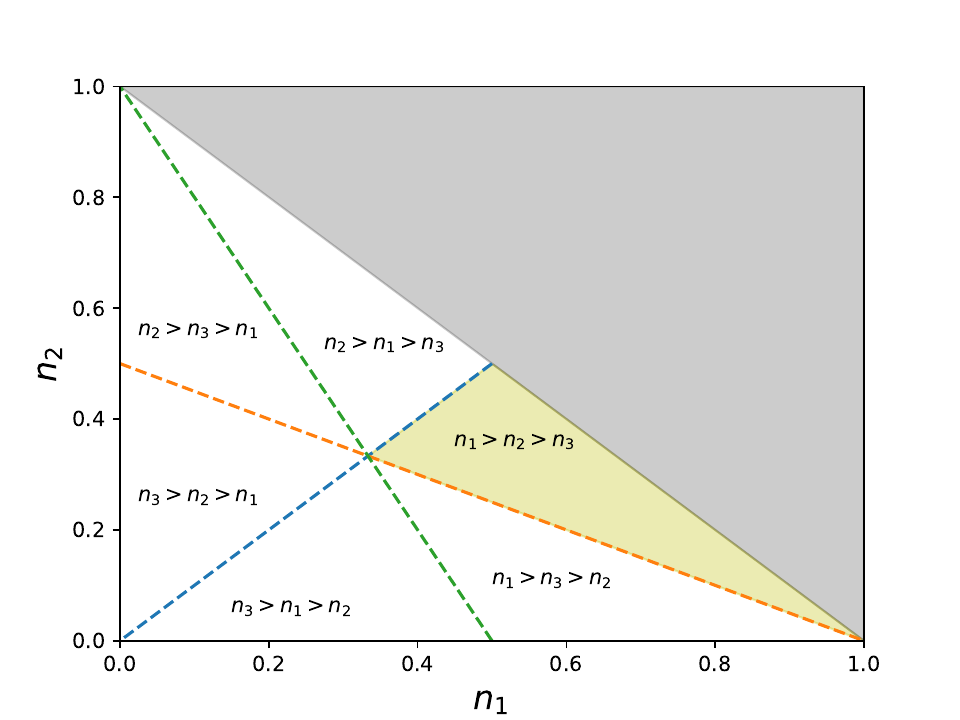}
	\includegraphics[width=.48\linewidth]{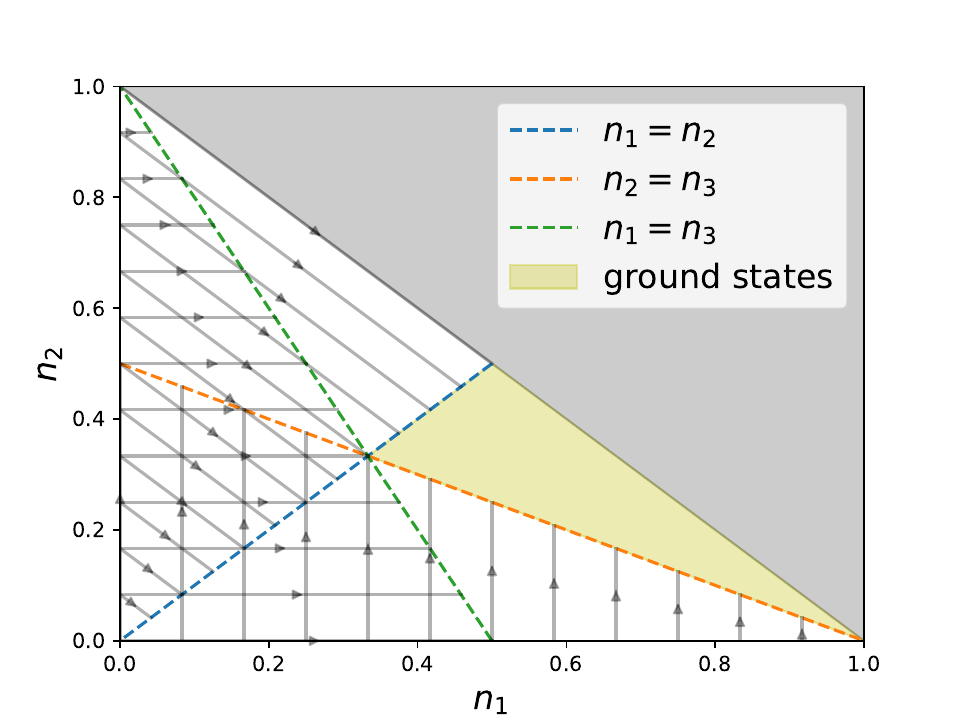}
	\caption{Left: the regions of state space corresponding to the six possible orderings of the three populations. Right: The allowable non-annealing trajectories through state space at each point.}
	\label{fig:stateSpace}
\end{figure*}

In Refs.~\cite{Hay2015} and \cite{Hay2017}, Hay, Schiff and Fisch approached the problem of calculating the \textit{maximum} accessible free energy in discrete systems -- that is, identifying the minimum-energy accessible ground state. 
In particular, Ref.~\cite{Hay2015} describes five primary findings for the $N = 3$ discrete system. 
To briefly paraphrase (and using numbering to match the original paper): 
\begin{enumerate}
\item For any given initial population values, it is possible to identify a finite number of accessible states whose associated energies could be extremal. 
In order to calculate the maximum accessible free energy, it suffices to identify these states and find the lowest-energy state among them. 
\item Candidates for the extremal states are always reachable within $N \text{ choose } 2$ averaging operations (that is, for $N = 3$, 3 operations). 
\item For any given initial conditions, there are ultimately seven candidates among the accessible states which may be extremal (including the initial state). Depending on the energy values assigned to each of the three states, any of these seven can be extremal. 
\item Allowing partial relaxation operations (partial mixing, as opposed to full averaging of a pair of populations) does not change the maximal energy that can be extracted from the system. 
\item Allowing steps that increase the energy instead of decreasing it (so-called annealing operations) does not change the maximal energy that can be extracted. 
\end{enumerate}
As it turns out, only result $4$ continues to hold when considering the problem of identifying the maximum-energy accessible ground state rather than the minimum-energy state. 

In some ways, this might seem surprising. 
Hay, Schiff, and Fisch's results were formulated in terms of the extremal accessible energies, not necessarily the minimum-energy states. 
There are two things which prevent most of their results from being directly applicable to the maximum-energy ground state problem. 
First: the highest-energy accessible state is not, in general, a ground state. 
As a result, the maximum-energy accessible \textit{ground} state is very often not one of the seven extremal states identified in Ref.~\cite{Hay2015}. 
Second: Hay, Schiff, and Fisch described how to calculate the set of states that are accessible when annealing operations are allowed. 
This made sense in the paper's original context, since, as they showed, annealing operations are never needed to reach the lowest-energy states. 
However, as discussed in Section~\ref{sec:definition}, it is not physically appropriate to allow annealing operations for the maximum-energy ground state problem, so the space of states to search for the maximum-energy ground state should be more restrictive than the solution space described in Ref.~\cite{Hay2015}. 

With that in mind, consider the question of identifying the maximum-energy accessible ground state when $N = 3$. 
In fact, it is possible to find a fairly compact solution to this problem by considering which operations are allowed for which starting states. 
This is probably easiest to understand graphically. 
Figure~\ref{fig:stateSpace} shows the space of possible populations $(n_1, n_2, n_3)$. 
As was noted in Ref.~\cite{Hay2015}, it is possible to represent this as a two-dimensional space by picking a normalization such that $n_1 + n_2 + n_3 = 1$ (in which case $n_3$ can be determined from the values of $n_1$ and $n_2$). 
Depending on the relative ordering of $n_1$, $n_2$, and $n_3$, different averaging operations are allowed in different regions of state space (according to the requirement that each averaging operation must decrease the energy of the system). 
The different orderings are shown in the left panel of Figure~\ref{fig:stateSpace}; the allowed trajectories for each region are shown in the right panel. 
An averaging operation consists of following one of the indicated trajectories to the $n_1 = n_2$, $n_2 = n_3$, or $n_1 = n_3$ line, depending on the averaging operation. 

The maximum-energy ground states can be read off of Figure~\ref{fig:stateSpace} region-by-region. 
If the initial state has $n_1 \geq n_2 \geq n_3$, then the system is already in a ground state, and the problem is trivial. 

If the initial state has $n_2 \geq n_1 \geq n_3$, then the only allowed operation is to average the first and second populations. 
The immediately results in the (only) accessible ground state:
\begin{gather*}
\def\arraystretch{1.2}
\begin{array}{|c|c|c|}
\hline
n_1 & n_2 & n_3\\ \hline
\end{array} \rightarrow 
\begin{array}{|c|c|c|}
\hline
\frac{1}{2}(n_1+n_2) & \frac{1}{2}(n_1 + n_2) & n_3\\ \hline
\end{array}
\; .
\end{gather*}
Similarly, if $n_1 \geq n_3 \geq n_2$, then the only allowed operation is to average the second and third populations, which again immediately brings the system to a ground state: 
\begin{gather*}
\def\arraystretch{1.2}
\begin{array}{|c|c|c|}
\hline
n_1 & n_2 & n_3\\ \hline
\end{array} \rightarrow 
\begin{array}{|c|c|c|}
\hline
n_1 & \frac{1}{2}(n_2+n_3) & \frac{1}{2}(n_2+n_3) \\ \hline
\end{array}
\; .
\end{gather*}

If $n_3 \geq n_2 \geq n_1$, then it is always possible to reach the ground state $(1/3, 1/3, 1/3)$, or at least to approach it arbitrarily closely. 
This can be done by alternating between averaging the populations of states 1 and 2 and averaging the populations of states 2 and 3. 
These two averaging operations, performed one after the other $k$ times, is equivalent to the mapping 
\begin{gather}
\begin{pmatrix}
n_1 \\ n_2 \\ n_3
\end{pmatrix} 
\rightarrow
\mathcal{A}^k
\begin{pmatrix}
n_1 \\ n_2 \\ n_3
\end{pmatrix} 
\end{gather}
with 
\begin{gather}
\mathcal{A} \doteq \begin{pmatrix}
1/2 & 1/2 & 0 \\
1/4 & 1/4 & 1/2 \\
1/4 & 1/4 & 1/2
\end{pmatrix} , 
\end{gather}
and for any $k \in \mathbb{N}$ it can be shown that 
\begin{gather}
\mathcal{A}^k
= 
\frac{1}{3}
\begin{pmatrix}
1 & 1 & 1 \\ 
1 & 1 & 1 \\
1 & 1 & 1
\end{pmatrix}
+ 
\frac{4^{-k}}{3}
\begin{pmatrix}
2 & 2 & -4 \\
-1 & -1 & 2 \\
-1 & -1 & 2
\end{pmatrix} . 
\end{gather}
As such, the system eventually converges to $(1/3, 1/3, 1/3)$ as $k \rightarrow \infty$. 
Moreover, each of these operations releases energy. 
To see this, note that if the system starts with $n_1 \leq n_2 \leq n_3$, averaging either the first and second or the second and third populations must release energy (or do nothing), and that neither of these operations will change the ordering of the three states' populations. 

This leaves two remaining cases: $n_2 > n_3 > n_1$ and $n_3 > n_1 > n_2$. 
Consider the former of these two. 
If $n_2 > n_3 > n_1$ and $(n_1 + n_2) / 2 \leq n_3$, then averaging the first and second populations gets the system to the boundary of the region discussed in the previous case, in which an alternating sequence of averaging operations between the first and second and second and third populations leads the system arbitrarily close to $(1/3,1/3,1/3)$. 
This is the highest-energy possible ground state, so it must be the optimal choice. 
On the other hand, if $(n_1 + n_2) / 2 > n_3$, then there are two possible allowed sequences of moves. 
Either the first and second can be averaged, leading to a ground state: 
\begin{gather*}
\def\arraystretch{1.2}
\begin{array}{|c|c|c|}
\hline
n_1 & n_2 & n_3\\ \hline
\end{array} \rightarrow 
\begin{array}{|c|c|c|}
\hline
\frac{1}{2}(n_1+n_2) & \frac{1}{2}(n_1 + n_2) & n_3\\ \hline
\end{array}
%\; .
\end{gather*}
or the first and third can be averaged, after which the only allowed operation is to average the first and second, leading to a ground state: 
\begin{align*}
\def\arraystretch{1.2}
&\begin{array}{|c|c|c|}
\hline
n_1 & n_2 & n_3\\ \hline
\end{array} \\
&\hspace{5 pt}\rightarrow 
\def\arraystretch{1.2}
\begin{array}{|c|c|c|}
\hline
\frac{1}{2}(n_1+n_3) & n_2 & \frac{1}{2}(n_1 + n_3) \\ \hline
\end{array} \\
&\hspace{5 pt}\rightarrow 
\def\arraystretch{1.2}
\begin{array}{|c|c|c|}
\hline
\frac{1}{4} n_1 + \frac{1}{2} n_2 + \frac{1}{4} n_3 & \frac{1}{4} n_1 + \frac{1}{2} n_2 + \frac{1}{4} n_3 & \frac{1}{2} n_1 + \frac{1}{2} n_3 \\ \hline
\end{array}
\; .
\end{align*}
The first of these two possible sequence always leads to a higher final energy, so it is the optimal choice. 
Intuitively, one can see this from Figure~\ref{fig:stateSpace}: moving horizontally in state space before moving diagonally down leads to a final ground state with a lower energy than would be reached by moving diagonally down from the initial position. 

The argument for the final region of initial state-space, in which $n_3 > n_1 > n_2$, is essentially the same. 
If $(n_2 + n_3) / 2 \geq n_1$, then averaging the second and third populations leads to the boundary of the region in which the highest-energy ground state, $(1/3, 1/3, 1/3)$, is reachable. 
If, on the other hand, $(n_2 + n_3) / 2 < n_1$, then it turns out always to be favorable to average the second and third populations, which immediately leads to a ground state: 
\begin{gather*}
\def\arraystretch{1.2}
\begin{array}{|c|c|c|}
\hline
n_1 & n_2 & n_3\\ \hline
\end{array} \rightarrow 
\begin{array}{|c|c|c|}
\hline
n_1 & \frac{1}{2}(n_2+n_3) & \frac{1}{2}(n_2 + n_3) \\ \hline
\end{array}
\; .
\end{gather*}
This is sufficient to specify, for any given initial state, the sequence of allowed operations that leads to the highest-energy possible ground state. 

At this point, it is possible to see in which ways the maximum-energy ground state problem differs from the minimum-energy ground state problem. 
Consider the five conclusions on the latter problem in Ref.~\cite{Hay2015}, listed at the beginning of this section. 

The first and third appear not to apply in the same way to the maximum-energy ground state problem; rather than identifying a finite set of candidate sequences and checking each, the solution presented here simply specifies directly which trajectory through state space is optimal, depending on the initial populations. 
However, although this was not the approach taken by Hay, Schiff, and Fisch, this kind of explicit case-by-case solution is also possible for the minimum-energy ground state problem. 
This is described in Appendix~\ref{appendix:threeBox}. 

The second conclusion from Ref.~\cite{Hay2015} (that the optimal ground state is always accessible within three operations) is entirely untrue for the maximum-energy ground state problem. 
In cases where the highest-energy accessible ground state is $(1/3,1/3,1/3)$, this optimal state is sometimes accessible only in the limit of an infinite number of operations. 
For example, the initial state $(0,1/4,3/4)$ can only lead to populations of the form $A/2^B$ for positive integers $A$ and $B$ for any finite number of steps; therefore it cannot reach $(1/3,1/3,1/3)$ in finite steps, but it is shown above that it can approach that ground state arbitrarily closely. 
The fifth conclusion (that allowing or prohibiting annealing operations does not change the optimal accessible state) also does not continue to hold for the present problem; this is discussed in Section~\ref{sec:definition}. 

The remaining, fourth conclusion -- that the optimal ground state is the same whether or not partial mixing operations are allowed -- is the only one that continues to hold for the $N=3$ maximum-energy ground state problem. 
``Partial relaxation" refers to any operation of the form 
\begin{gather}
n_i \rightarrow (1-\gamma) n_i + \gamma n_j \\
n_j \rightarrow \gamma n_i + (1-\gamma) n_j
\end{gather}
for $0 < \gamma < 1/2$ (rather than ``full mixing," where $\gamma = 1/2$). 
It is easiest to see this by inspecting the trajectories in Figure~\ref{fig:stateSpace}. 
A partial mixing operation would still have to follow one of the marked trajectories, but unlike a full mixing operation, it would not have to follow a given trajectory line to one of the $n_i = n_j$ lines. 
For initial conditions where $n_2 > n_1 > n_3$ and $n_1 > n_3 > n_2$, there is only one allowed pair of populations to average anyway. 
For initial conditions where $n_3 > n_2 > n_1$, full mixing operations can already reach the highest-possible-energy ground state $(1/3,1/3,1/3)$, so it is clear that no other operations could do better. 
The cases in which $n_2 > n_3 > n_1$ or $n_1 > n_3 > n_2$ are less trivial, but still clear from the figure: the operations which average the first and third populations are never favorable, whether they are complete or partial, so the optimal first move is always to fully mix the first and second populations (if $n_2 > n_3 > n_1$) or the second and third (if $n_1 > n_3 > n_2$). 

Some of these conclusions (for instance, the role of annealing operations) hold for all $N$. 
Others (like the effects of partial mixing) seem likely to continue to hold when $N > 3$, but we have not proved them here for general $N$. 

Note that there is no case in which reaching the optimal ground state requires mixing the first and third populations. 
The optimal sequences only ever require that neighboring states be mixed. 
We have proven this for the three-state discrete case, but we conjecture that it is true for all $N$. 

\section{Continuous Example: The Bump-on-Tail Distribution} \label{sec:continuous}

Consider a bump-on-tail distribution $f(v)$. 
$f(v)$ is monotonically decreasing until it hits a local minimum, then monotonically increasing until it arrives at a local maximum, and thenceforth monotonically decreasing. 
We will assume that $f(v=0)$ exceeds this local maximum, and that the global minimum happens as $v \rightarrow \infty$; these assumptions are not necessary for what follows, but they are convenient. 
Let the energy of a particle with velocity $v$ be given by $\varepsilon(v) = m v^2 / 2$ for some mass $m$. 

The ``quasilinear plateau" is constructed by finding velocities $v_1$ and $v_2$ such that we can construct a flattened function $\bar{f}(v)$ as follows: 
\begin{gather}
\bar{f}(v) \doteq \begin{cases}
f(v) & v < v_1 \text{ or } v > v_2 \\
h & v_1 \leq v \leq v_2, 
\end{cases}
\end{gather}
with 
\begin{gather}
h \doteq \frac{1}{v_2 - v_1} \int_{v_1}^{v_2} f(v) \, \D v . 
\end{gather}
For a bump-on-tail distribution, $h$ is chosen as the unique value for which $f(v_1) = f(v_2) = h$. 
This section will demonstrate that $\bar{f}(v)$ is the maximum-energy accessible ground state for the bump-on-tail distribution. 

In addition to $v_1$ and $v_2$, there are two other important values of $v$ to note: first, $v_0$, the minimal value of $v$ at which $f(v_0)$ attains the same value as the bump's local maximum of $f$; and second, $v_3$, the maximal value of $v$ at which $f(v_3)$ attains the same value s the bump's local minimum of $f$. Note that for this starting distribution, $v_0 < v_1 < v_2 < v_3$. 

The plateau distribution $\bar{f}(v)$ is accessible through diffusive operations. 
One can pair intervals within $[v_1, v_2]$ over which $f(v) > h$ with those by which $f(v) < h$ and successively exchanging particles between them until they converge to $f(v) = h$. 
This would not require annealing, since the intervals within $[v_1, v_2]$ for which $f(v) > h$ all occur at lower $v$ than those for which $f(v) < h$. 

There can be no exchange involving $v < v_0$ or $v > v_3$; any exchange involving these intervals would require annealing. 
Moreover, it is clear that within $[v_1, v_2]$, it is impossible to do better than a flat distribution. 
As such, the only scenario in which one might imagine accessing a higher-energy ground state than the plateau is if there were some exchanges involving the intervals $[v_0,v_1]$ or $[v_2,v_3]$. 

Any non-annealing exchange involving $[v_0, v_1]$ must transfer population into this region. 
Any non-annealing exchange involving $[v_2, v_3]$ must transfer population out of this region. 
This follows from the fact that $f(v)$ is monotonically decreasing for $v < v_1$ and $v > v_2$. 
Therefore, a higher-energy ground state would have to involve exchanges that move some total population $F_L \geq 0$ to $[v_0, v_1]$ from $[v_1,v_2]$ and some total population $F_R \geq 0$ to $[v_1, v_2]$ from $[v_2, v_3]$ (with at least one of $F_L$ and $F_R$ being nonzero). 
The resulting ground state would have an energy bounded above by the case in which $f(v)$ could still be flattened between $v_1$ and $v_2$. 
But then any $F_L > 0$ must lower the distribution's total energy in $[v_1, v_2]$ by more than it increased the energy in $[v_0, v_1]$, and any $F_R > 0$ must lower the distribution's energy in $[v_2, v_3]$ by more than it increased the energy in $[v_1, v_2]$. 
In other words, the exchanges involving the regions $[v_0, v_1]$ and $[v_2, v_3]$ never lead to a ground state with an energy higher than that of the quasilinear plateau. 

\begin{figure}
	\centering
	\includegraphics[width=.99\linewidth]{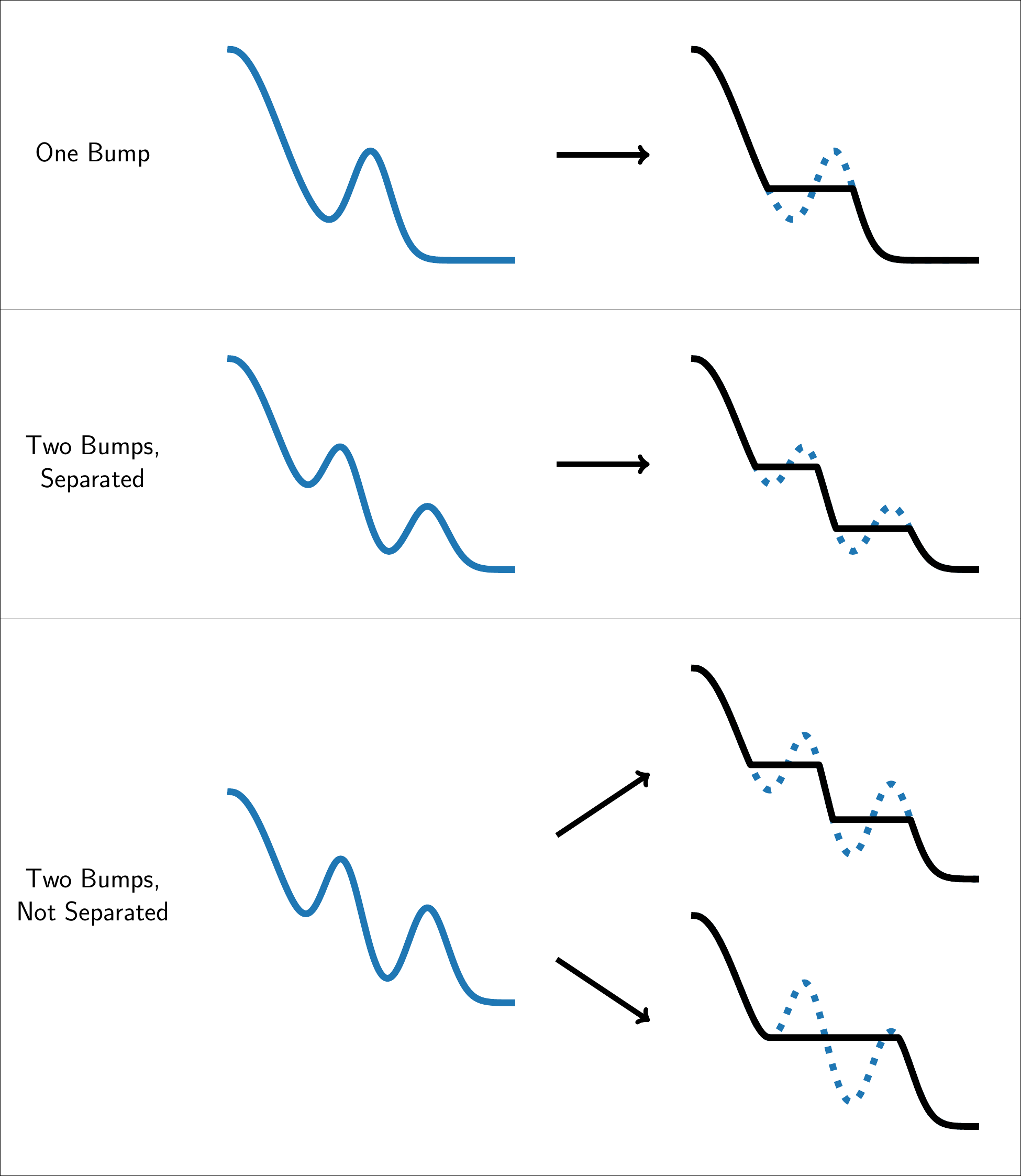}
	\caption{This cartoon shows how a starting distribution with more than one bump on its tail can lead to multiple ``plateau-like" ground states, if the bumps are not sufficiently separated (in particular, if the local minimum of $f$ associated with one bump does not exceed the local maximum associated with the other). The one-bump case is the paradigmatic ``bump-on-tail" distribution with the classic, textbook plateau solution that is adjusted in height to conserve particles by matching the area below the plateau line with that above it. } 
	\label{fig:twoBumpCartoon}
\end{figure}

This is enough to determine the maximum-energy accessible ground state for one class of distribution functions (albeit an important one). 
A logical next case to consider is a distribution with multiple bumps on its tail. 
The generalization is straightforward in cases where the two bumps are sufficiently separated. 
In particular, if the local minimum of $f(v)$ for the lower-$v$ bump exceeds the local maximum of $f(v)$ for the higher-$v$ bump, then the two plateau regions cannot interact without annealing operations and there is a unique two-plateau solution which must be the maximum-energy accessible ground state. 
Things are more complicated if the two bumps are not separated in this way. 
This is illustrated in Figure~\ref{fig:twoBumpCartoon}; it is possible for there to be multiple ``plateau-like" ground states. 
If it were true that the maximum-energy ground state is always plateau-like, then this means that computing the maximum-energy ground state would still be a nontrivial search problem over some set of plateau-like candidate distributions. 
We conjecture that the maximum-energy ground state is, in fact, plateau-like for any sufficiently well-behaved smooth function, but we have not proved that this must be the case. 
If so, then it would follow that continuous maximum-energy ground states are accessible through local diffusion. 

\section{Conclusion} \label{sec:conclusion}

There is a class of problems in which a distribution of particles is rearranged via phase space diffusion. 
The most familiar example in plasma physics is the quasilinear diffusion that appears in the theory of weak turbulence. 
The same formalism also applies to other systems, including laser-stimulated emissions due to transitions between atomic states. 
In these problems, it is desirable to understand the behavior of a distribution under the influence of diffusive operations. 

Diffusive operations are known to be able to map a given initial system to a spectrum of different ground states. 
Previous work has always focused on bounding the final energy of that spectrum from below -- that is, determining the maximum energy that could be released \cite{Fisch1992, Hay2015, Hay2017, Kolmes2020ConstrainedDiffusion, Kolmes2020Gardner}. 
This paper has argued that the upper bound of the ground-state energy spectrum (which sets the lower bound for how much energy could be released) is comparably important, especially for the purposes of understanding uncontrolled instabilities. 

Moreover, this paper has identified the maximum-energy ground state in certain cases. 
In a discrete phase space with $N = 3$ elements, the maximum-energy ground state can be determined by carefully considering the allowable mixing operations at every point in state space. 
In a continuous phase space, it turns out that the quasilinear plateau is the maximum-energy diffusively accessible ground state for the bump-on-tail distribution. 
This means that the maximum-energy ground state can be understood as a natural generalization of the quasilinear plateau for general distributions. 
The quasilinear plateau is a paradigm of recurring interest in this literature because it is the best-known and most intuitive example of what a diffusively accessible ground state could look like; here we identify precisely where on the spectrum of ground state energies it falls and find that it is extremal. 

These examples lead us to make two conjectures: 
\begin{enumerate}
	\item For a large class of smooth initial distributions, the maximum-energy ground
	states are plateau-like, in the sense that they consist of segments of distribution that are fully flattened and segments which are not modified from their initial forms.
	\item In both the discrete and continuous cases, only local mixing operations are
	necessary in order to reach the maximum-energy ground states.
\end{enumerate}
It would certainly be interesting to know whether either of these conjectures are true, but neither of these conjectures is proved here. However, based upon known examples, neither is disproved and both appear to be plausible. 
More generally, this paper serves to introduce a previously overlooked problem that helps to characterize the entire spectrum of possible ground state energies, rather than focusing on their lower bound alone.

\acknowledgements
This work was supported by US DOE DE-SC0016072.

\providecommand{\noopsort}[1]{}\providecommand{\singleletter}[1]{#1}%

\appendix
\section{Exact Solutions to the $N=2$ and $N=3$ Gardner Restacking and Maximum-Energy-Release Diffusive Exchange Problems} \label{appendix:threeBox}

\subsection{Introduction to the Problem}
Section~\ref{sec:threeBox} presents an explicit solution for the problem of finding the maximum-energy accessible ground state for a discrete system with three states. 
This appendix will describe the corresponding solutions for the minimum-energy ground state accessible through diffusion, as well as describing the Gardner restacking solutions. 

The objective of the minimum-energy ground state problem is, given a collection of $N$ initial populations $\{n_i\}$ and corresponding energies $\varepsilon_i$, to find the sequence of pairwise averaging operations on the $\{n_i\}$ that will minimize the energy 
\begin{gather}
W = \sum_i \varepsilon_i n_i . 
\end{gather}
It is typically convenient to use a convention in which the cells are arranged in increasing order of energy, that is, $\varepsilon_i \leq \varepsilon_j$ $\forall i < j$. 
The difference between the initial energy $W$ and the minimal final energy is $\Delta W_\text{D}$, the (maximal) diffusive free energy. 
In general, finding the minimal final $W$ is a computationally intensive problem. When there are $N$ discrete elements, Hay, Schiff, and Fisch showed \cite{Hay2015} that the search space of possible optimal sequences of exchanges has an $\mathcal{O}(N^{N^2})$ upper bound. 
However, the problem can be tractable -- even without computer assistance -- for sufficiently small $N$. This appendix will present a general solution for the diffusive-exchange problems for $N = 2$ and $N = 3$. 

This appendix is adapted from material that appeared in one of the authors' doctoral dissertations \cite{KolmesThesis}; we reproduce it here because it has not previously appeared in the archival literature. 

\subsection{The $N=2$ Problem}

Consider a system with initial populations $n_1$ and $n_2$ and corresponding energies $\varepsilon_1$ and $\varepsilon_2$. Assume without loss of generality that $\varepsilon_1 \leq \varepsilon_2$. If $n_1 \geq n_2$, then the system is already in a ground state, and $\WG = \WD = 0$. If, on the other hand, $n_1 < n_2$, then the ground state can be reached as follows: 
\begin{gather}
\def\arraystretch{1.2}
\begin{array}{|c|c|}
\hline
n_1 & n_2 \\ \hline
\end{array} \rightarrow
\def\arraystretch{1.2} 
\begin{array}{|c|c|}
\hline
\frac{1}{2} (n_1 + n_2) & \frac{1}{2} (n_1 + n_2) \\ \hline
\end{array}
\; .
\end{gather}
Gardner restacking instead does the following: 
\begin{gather}
\def\arraystretch{1.2}
\begin{array}{|c|c|}
\hline
n_1 & n_2 \\ \hline
\end{array} \rightarrow
\def\arraystretch{1.2} 
\begin{array}{|c|c|}
\hline
n_2 & n_1 \\ \hline
\end{array}
\; .
\end{gather}
In this case, $\WG = (\varepsilon_2 - \varepsilon_1) (n_2 - n_1)$ and $\WD = (\varepsilon_2 - \varepsilon_1) (n_2 - n_1) / 2$. In fact, in either of the two possible cases, it turns out that $\WG = 2 \WD$. 

\subsection{The $N=3$ Problem}

Consider a three-state system with initial populations $n_1$, $n_2$, and $n_3$ and corresponding energies $\varepsilon_1 \leq \varepsilon_2 \leq \varepsilon_3$. For the purposes of this problem, only the differences between populations and energies matter, so in fact this problem depends only on the following four parameters: 
\begin{align}
&\Delta_1 \doteq n_2 - n_1 \\
&\Delta_2 \doteq n_3 - n_2 \\
&\alpha \doteq \varepsilon_2 - \varepsilon_1 \\
&\beta \doteq \varepsilon_3 - \varepsilon_2. 
\end{align}
$\Delta_1$ and $\Delta_2$ are defined in terms of the initial values of $n_1$, $n_2$, and $n_3$. The arguments that follow will rely on the fact -- proven in Ref.~\cite{Hay2015} -- that the optimal sequence of diffusive exchange operations for minimizing the final ground-state energy never needs to include so-called ``annealing" operations; in other words, it is safe to assume that every averaging step in the optimal sequence must decrease $W$. Ref.~\cite{Hay2015} also showed that the optimal sequence of exchanges on $N$ cells can include at most ($N$ choose 2) transformations. For $N = 3$, this means that the optimal sequence will never involve more than 3 transformations. 

In order to calculate $\WD$, we will consider each of the possible orderings of the initial values of $n_1$, $n_2$, and $n_3$. This results in a total of six cases to consider. Note that there are corner cases which fit into more than one of the cases below (for instance, if two of the populations are initially equal). 

\textbf{Case 1: $n_1 \geq n_2 \geq n_3$.} In this case, there is no diffusive exchange operation (nor any restacking exchange operation) that can reduce the initial energy. The system is already in its ground state: 
\begin{align}
&\WG = 0 \\
&\WD = 0. 
\end{align}

\textbf{Case 2: $n_1 \geq n_3 \geq n_2$.} In this case, it is never profitable to perform an exchange involving $n_1$, so there is only one candidate for the optimal sequence: to average $n_2$ and $n_3$: 
\begin{gather*}
\def\arraystretch{1.2}
\begin{array}{|c|c|c|}
\hline
n_1 & n_2 & n_3\\ \hline
\end{array} \rightarrow 
\def\arraystretch{1.2}
\begin{array}{|c|c|c|}
\hline
n_1 & \frac{1}{2}(n_2+n_3) & \frac{1}{2}(n_2 + n_3) \\ \hline
\end{array}
\; .
\end{gather*}
The result is a ground state with energy less than the starting energy. Of course, the Gardner free energy can be found by performing a restacking exchange on the same two elements: 
\begin{gather*}
\def\arraystretch{1.2}
\begin{array}{|c|c|c|}
\hline
n_1 & n_2 & n_3\\ \hline
\end{array} \rightarrow 
\def\arraystretch{1.2}
\begin{array}{|c|c|c|}
\hline
n_1 & n_3 & n_2\\ \hline
\end{array}
\; .
\end{gather*}
As a result, 
\begin{align}
&\WG = (\varepsilon_3 - \varepsilon_2) (n_3 - n_2) = \beta \Delta_2 \\
&\WD = \frac{1}{2} \, \beta \Delta_2. 
\end{align}

\textbf{Case 3: $n_2 \geq n_1 \geq n_3$.} This case is much the same as Case 2, except that the only possible exchange is between $n_1$ and $n_2$, after which the system is in its minimal-energy ground state. Therefore, 
\begin{align}
&\WG = (\varepsilon_2 - \varepsilon_1) (n_2 - n_1) = \alpha \Delta_1 \\
&\WD = \frac{1}{2} \, \alpha \Delta_1. 
\end{align}

\textbf{Case 4: $n_2 \geq n_3 \geq n_1$.} This time, there are two possible diffusive ``starting moves": to average $n_1$ and $n_2$ or to average $n_1$ and $n_3$. Since there are now multiple ground states accessible without any annealing operations, it is helpful to introduce the notation that $T_{ij}$ is the operation of averaging cells $i$ and $j$, and that $T_{ij} T_{\ell k}$ is the operation of averaging cells $\ell$ and $k$, followed by the operation of averaging cells $i$ and $j$. 

The optimal sequence must begin with either $T_{12}$ or $T_{13}$. If it begins with $T_{12}$, then either $2n_3 \leq n_1 + n_2$ (in which case the system is in a ground state) or $2 n_3 < n_1 + n_2$, in which case the only remaining sequence of less than four total non-annealing exchanges leading to a ground state is $T_{23} T_{13} T_{12}$. 
On the other hand, if the sequence begins with $T_{13}$, then the only choice for the next operation is $T_{12}$, which brings the system to a ground state. 

As a result, there are three viable candidates for the optimal sequence: $T_{12}$, $T_{23} T_{13} T_{12}$, and $T_{12} T_{13}$. The resulting ground states are as follows: 
\begin{widetext}
\begin{align}
T_{12}: & \; \def\arraystretch{1.2} \begin{array}{|c|c|c|}
\hline
\frac{1}{2} n_1 + \frac{1}{2} n_2 & \frac{1}{2} n_1 + \frac{1}{2} n_2 & n_3\\ \hline
\end{array} \\
T_{23} T_{13} T_{12}: & \; \def\arraystretch{1.2} \begin{array}{|c|c|c|}
\hline
\frac{1}{4} n_1 + \frac{1}{4} n_2 + \frac{1}{2} n_3 & \frac{3}{8} n_1 + \frac{3}{8} n_2 + \frac{1}{4} n_3 & \frac{3}{8} n_1 + \frac{3}{8} n_2 + \frac{1}{4} n_3 \\ \hline
\end{array} \\
T_{12} T_{13}: & \; \def\arraystretch{1.2} \begin{array}{|c|c|c|}
\hline
\frac{1}{4} n_1 + \frac{1}{2} n_2 + \frac{1}{4} n_3 & \frac{1}{4} n_1 + \frac{1}{2} n_2 + \frac{1}{4} n_3 & \frac{1}{2} n_1 + \frac{1}{2} n_3 \\ \hline
\end{array} \, . 
\end{align}
\end{widetext}
After $T_{12}$, the total released energy is 
\begin{gather}
\Delta W_{T_{12}} = \frac{1}{2} \, \alpha \Delta_1. 
\end{gather}
After $T_{23} T_{13} T_{12}$, the released energy is 
\begin{align}
&\Delta W_{T_{23} T_{13} T_{12}} \nonumber \\
&\hspace{10 pt}= \alpha \bigg( \frac{3}{8} \Delta_1 - \frac{1}{4} \Delta_2 \bigg) + (\alpha + \beta) \bigg( \frac{3}{8} \Delta_1 + \frac{3}{4} \Delta_2 \bigg) \\
&\hspace{10 pt}= \alpha \bigg( \frac{3}{4} \Delta_1 + \frac{1}{2} \Delta_2 \bigg) + \beta \bigg( \frac{3}{8} \Delta_1 + \frac{3}{4} \Delta_2 \bigg) . 
\end{align}
Finally, after $T_{12} T_{13}$, the released energy is 
\begin{align}
&\Delta W_{T_{12} T_{13}} \nonumber \\
&\hspace{10 pt}= \alpha \bigg( \frac{1}{4} \Delta_1 - \frac{1}{4} \Delta_2 \bigg)  + (\alpha + \beta) \bigg( \frac{1}{2} \Delta_1 + \frac{1}{2} \Delta_2 \bigg) \\ 
&\hspace{10 pt}= \alpha \bigg( \frac{3}{4} \Delta_1 + \frac{1}{4} \Delta_2 \bigg) + \beta \bigg( \frac{1}{2} \Delta_1 + \frac{1}{2} \Delta_2 \bigg) . 
\end{align}
Note that in this scenario, $\Delta_1 \geq 0$, $\Delta_2 \leq 0$, and $\Delta_1 + \Delta_2 \geq 0$. Perhaps surprisingly, $\Delta W_{T_{12} T_{13}}$ is always the largest of the three, regardless of the relative sizes of the differences in the three cells' initial populations or energies. 

The Gardner free energy can be found, as usual, simply by reordering the $\{n_i\}$ to put the higher populations in the lower-energy cells. As such, 
\begin{align}
&\WG = \alpha \Delta_1 + \beta (\Delta_1 + \Delta_2) \\
&\WD = \alpha \bigg( \frac{3}{4} \Delta_1 + \frac{1}{4} \Delta_2 \bigg) + \beta \bigg( \frac{1}{2} \Delta_1 + \frac{1}{2} \Delta_2 \bigg) . 
\end{align}

\textbf{Case 5: $n_3 \geq n_1 \geq n_2$.} This case has much in common with the previous one. Again, there are two possible starting exchanges for a candidate optimal sequence. This time, they are $T_{13}$ and $T_{23}$. 
If the candidate sequence begins with $T_{23}$, then the system is already in a ground state if $2 n_1 \geq n_2 + n_3$. Otherwise, the only sequence of moves that is short enough to be a candidate, that results in a ground state, and that does not include annealing moves is $T_{12} T_{13} T_{23}$. 

If the sequence instead begins with $T_{13}$, then it is not immediately in a ground state, since $n_2$ is the smallest of the three initial populations (not counting the trivial corner case in which $n_1 = n_2 = n_3$). Then the only remaining move to follow $T_{13}$ is $T_{23}$, which results in a ground state. The three candidate ground states are as follows: 
\begin{widetext}
\begin{align}
T_{23}: & \; \def\arraystretch{1.2} \begin{array}{|c|c|c|}
\hline
n_1 & \frac{1}{2} n_2 + \frac{1}{2} n_3 & \frac{1}{2} n_2 + \frac{1}{2} n_3 \\ \hline
\end{array} \\
T_{12} T_{13} T_{23}: & \; \def\arraystretch{1.2} \begin{array}{|c|c|c|}
\hline
\frac{1}{4} n_1 + \frac{3}{8} n_2 + \frac{3}{8} n_3 & \frac{1}{4} n_1 + \frac{3}{8} n_2 + \frac{3}{8} n_3 & \frac{1}{2} n_1 + \frac{1}{4} n_2 + \frac{1}{4} n_3 \\ \hline
\end{array} \\
T_{23} T_{13}: & \; \def\arraystretch{1.2} \begin{array}{|c|c|c|}
\hline
\frac{1}{2} n_1 + \frac{1}{2} n_3 & \frac{1}{4} n_1 + \frac{1}{2} n_2 + \frac{1}{4} n_3 & \frac{1}{4} n_1 + \frac{1}{2} n_2 + \frac{1}{4} n_3 \\ \hline
\end{array} \, .
\end{align}
\end{widetext}
The corresponding released energy for $T_{23}$ is 
\begin{align}
\Delta W_{T_{23}} &=  \frac{1}{2} \, \beta \Delta_2 . 
\end{align}
The released energy for $T_{12} T_{13} T_{23}$ is 
\begin{align}
\Delta W_{T_{12} T_{13} T_{23}} &= \alpha \bigg( \frac{1}{4} \Delta_1 - \frac{3}{8} \Delta_2 \bigg) + (\alpha+\beta) \bigg( \frac{1}{2} \Delta_1 + \frac{3}{4} \Delta_2 \bigg) \\
&= \alpha \bigg( \frac{3}{4} \Delta_1 + \frac{3}{8} \Delta_2 \bigg) + \beta \bigg( \frac{1}{2} \Delta_1 + \frac{3}{4} \Delta_2 \bigg) . 
\end{align}
The released energy corresponding to $T_{23} T_{13}$ is 
\begin{align}
&\Delta W_{T_{23} T_{13}} \nonumber \\
&\hspace{10 pt}= \alpha \bigg( \frac{1}{4} \Delta_1 - \frac{1}{4} \Delta_2 \bigg) + (\alpha + \beta) \bigg( \frac{1}{4} \Delta_1 + \frac{3}{4} \Delta_2 \bigg) \\
&\hspace{10 pt}= \alpha \bigg( \frac{1}{2} \Delta_1 + \frac{1}{2} \Delta_2 \bigg) + \beta \bigg( \frac{1}{4} \Delta_1 + \frac{3}{4} \Delta_2 \bigg) . 
\end{align}
Noting that in the present case $\Delta_1 \leq 0$, $\Delta_2 \geq 0$, and $\Delta_1 + \Delta_2 \geq 0$, the largest of these is always $\Delta W_{T_{23} T_{13}}$. Then, computing the Gardner free energy in the usual way, 
\begin{align}
&\WG = \alpha (\Delta_1 + \Delta_2) + \beta \Delta_2 \\
&\WD = \alpha \bigg( \frac{1}{2} \Delta_1 + \frac{1}{2} \Delta_2 \bigg) + \beta \bigg( \frac{1}{4} \Delta_1 + \frac{3}{4} \Delta_2 \bigg) . 
\end{align}

\textbf{Case 6: $n_3 \geq n_2 \geq n_1$.} In this final case, the initial populations are arranged in such a way that averaging any of the three pairs of cells (1 and 2, 2 and 3, or 1 and 3) would release energy. 

First consider a candidate sequence beginning with $T_{12}$. $T_{12}$ does not immediately result in a ground state. The two possibilities for a subsequent transformation are $T_{13}$ and $T_{23}$. After $T_{12}$ and $T_{13}$, there is one possible move that leads to a ground state: $T_{23}$. However, after $T_{12}$ and $T_{23}$, there is no sequence of three or less total moves that can lead to a ground state. So the only viable candidate beginning with $T_{12}$ is $T_{23} T_{13} T_{12}$. 

Now consider a sequence beginning with $T_{13}$. This first move does not immediately produce a ground state, and must be followed with either $T_{12}$ or $T_{23}$. In either case, either it reaches a ground state in two moves or it cannot reach one in three or less. As a result, there are two candidate sequences beginning with $T_{13}$: that is, $T_{12} T_{13}$ and $T_{23} T_{13}$. 

Lastly, consider a sequence beginning with $T_{23}$. This sub-case works out in much the same way as the candidate sequences beginning with $T_{12}$. After enumerating the possibilities, there is only one that reaches a possible ground state within three exchanges without annealing. It is $T_{12} T_{13} T_{23}$. 

The four resulting candidates are: 
\begin{widetext}
\begin{align}
T_{23} T_{13} T_{12}: & \; \def\arraystretch{1.2} \begin{array}{|c|c|c|}
\hline
\frac{1}{4} n_1 + \frac{1}{4} n_2 + \frac{1}{2} n_2 & \frac{3}{8} n_1 + \frac{3}{8} n_2 + \frac{1}{4} n_3 & \frac{3}{8} n_1 + \frac{3}{8} n_2 + \frac{1}{4} n_3 \\ \hline \end{array} \\
T_{12} T_{13}: & \; \def\arraystretch{1.2} \begin{array}{|c|c|c|} 
\hline
\frac{1}{4} n_1 + \frac{1}{2} n_2 + \frac{1}{4} n_3 & \frac{1}{4} n_1 + \frac{1}{2} n_2 + \frac{1}{4} n_3 & \frac{1}{2} n_1 + \frac{1}{2} n_3 \\
\hline \end{array} \\
T_{23} T_{13}: & \; \def\arraystretch{1.2} \begin{array}{|c|c|c|}
\hline
\frac{1}{2} n_1 + \frac{1}{2} n_3 & \frac{1}{4} n_1 + \frac{1}{2} n_2 + \frac{1}{4} n_3 & \frac{1}{4} n_1 + \frac{1}{2} n_2 + \frac{1}{4} n_3 \\
\hline \end{array} \\
T_{12} T_{13} T_{23}: & \; \def\arraystretch{1.2} \begin{array}{|c|c|c|}
\hline 
\frac{1}{4} n_1 + \frac{3}{8} n_2 + \frac{3}{8} n_3 & \frac{1}{4} n_1 + \frac{3}{8} n_2 + \frac{3}{8} n_3 & \frac{1}{2} n_1 + \frac{1}{4} n_2 + \frac{1}{4} n_3 \\
\hline 
\end{array} \, .
\end{align}
\end{widetext}
Looking at these ground states, it is possible to see immediately that $T_{12} T_{13}$ can never release more energy than $T_{12} T_{13} T_{23}$, since the latter has a larger population differential between its first two cells and its third. Similarly, $T_{23} T_{13}$ can never release more energy than $T_{23} T_{13} T_{12}$. However, it is worthwhile to calculate the energies released by the remaining two candidates. $T_{23} T_{13} T_{12}$ releases 
\begin{align}
&\Delta W_{T_{23} T_{13} T_{12}} \nonumber \\
&\hspace{10 pt}= \alpha \bigg( \frac{3}{8} \Delta_1 - \frac{1}{4} \Delta_2 \bigg) + (\alpha + \beta) \bigg( \frac{3}{8} \Delta_1 + \frac{3}{4} \Delta_2 \bigg) \\
&\hspace{10 pt}= \alpha \bigg( \frac{3}{4} \Delta_1 + \frac{1}{2} \Delta_2 \bigg) + \beta \bigg( \frac{3}{8} \Delta_1 + \frac{3}{4} \Delta_2 \bigg) . 
\end{align}
%$T_{23} T_{13}$ releases
%\begin{align}
%\Delta W_{T_{23} T_{13}} &= \alpha \bigg( \frac{1}{4} \Delta_1 - \frac{1}{4} \Delta_2 \bigg) + (\alpha + \beta) \bigg( \frac{1}{4} \Delta_1 + \frac{3}{4} \Delta_2 \bigg) \\
%&= \alpha \bigg( \frac{1}{2} \Delta_1 + \frac{1}{2} \Delta_2 \bigg) + \beta \bigg( \frac{1}{4} \Delta_1 + \frac{3}{4} \Delta_2 \bigg) . 
%\end{align}
$T_{12} T_{13} T_{23}$ releases
\begin{align}
&\Delta W_{T_{12} T_{13} T_{23}} \nonumber \\
&\hspace{10 pt}= \alpha \bigg( \frac{1}{4} \Delta_1 - \frac{3}{8} \Delta_2 \bigg) + (\alpha + \beta) \bigg( \frac{1}{2} \Delta_1 + \frac{3}{4} \Delta_2 \bigg) \\
&\hspace{10 pt}= \alpha \bigg( \frac{3}{4} \Delta_1 + \frac{3}{8} \Delta_2 \bigg) + \beta \bigg( \frac{1}{2} \Delta_1 + \frac{3}{4} \Delta_2 \bigg). 
\end{align}
The difference between the two is 
\begin{gather}
\Delta W_{T_{12} T_{13} T_{23}} - \Delta W_{T_{23} T_{13} T_{12}} = \frac{1}{8} \bigg( \beta \Delta_1 - \alpha \Delta_2 \bigg). 
\end{gather}
In other words, $\Delta W_{T_{12} T_{13} T_{23}}$ is the larger of the two if $\beta \Delta_1 > \alpha \Delta_2$ and $\Delta W_{T_{23} T_{13} T_{12}}$ is the larger in the opposite case. Of course, they are equal if $\beta \Delta_1 = \alpha \Delta_2$. 

The conclusion is that in this case -- unlike any of the others -- $\WD$ depends on the relative sizes of the energy and population gaps in the starting populations (not just their signs). $\WG$, of course, follows from the same sorting argument as ever, and has no such ambiguity. The two free energies are: 
\begin{align}
\WG = (\alpha + \beta) ( \Delta_1 + \Delta_2) 
\end{align}
and 
\begin{widetext}
\begin{align}
\WD = \begin{cases}
\alpha \bigg( \dfrac{3}{4} \Delta_1 + \dfrac{1}{2} \Delta_2 \bigg) + \beta \bigg( \dfrac{3}{8} \Delta_1 + \dfrac{3}{4} \Delta_2 \bigg) & \text{if } \; \alpha \Delta_2 \geq \beta \Delta_1 \vspace{5 pt} \\ 
\alpha \bigg( \dfrac{3}{4} \Delta_1 + \dfrac{3}{8} \Delta_2 \bigg) + \beta \bigg( \dfrac{1}{2} \Delta_1 + \dfrac{3}{4} \Delta_2 \bigg) & \text{if } \; \alpha \Delta_2 < \beta \Delta_1. 
\end{cases}
\end{align}
\end{widetext}

\subsection{A Few Comments}

There are a few things to say about this analysis. First: if this is the amount of effort required to solve the diffusive problem with $N=3$, one can imagine how quickly the problem can spiral into computational intractability. The Gardner restacking problem is comparatively very easy. 

Second: if we look carefully at the difference between $\WG$ and $\WD$, we already find indications that $\WD$ might be getting closer to $\WG$ as $N$ gets larger. When $N = 2$, $\WD = \WG / 2$ in all (well, both) cases. When $N = 3$, $\WD \geq \WG / 2$ in each of the six cases. For example, consider Case 6. Depending on the values of $\alpha$, $\beta$, $\Delta_1$, and $\Delta_2$, $\WD$ can be anywhere from $(1/2) \WG$ to $(3/4) \WG$. This provides a hint of intuition for a result made rigorous in Ref.~\cite{Kolmes2020Gardner}: that in the continuous limit (where $N \rightarrow \infty$ and the variation between neighboring values is reasonably well-behaved) these two free energies become the same. 
	
\end{document}